\def\preprint{0} 

\if 1\preprint
\documentclass[sigconf,nonacm]{acmart}
\setcopyright{none}
\renewcommand\footnotetextcopyrightpermission[1]{}
\else
\documentclass[sigconf,nonacm]{acmart}
\fi

\usepackage{pdfpages}

\usepackage{listings}
\usepackage{color}

\definecolor{dkgreen}{rgb}{0,0.6,0}
\definecolor{gray}{rgb}{0.5,0.5,0.5}
\definecolor{mauve}{rgb}{0.58,0,0.82}
\definecolor{aureolin}{rgb}{0.99, 0.93, 0.0}
\definecolor{bananayellow}{rgb}{1.0, 0.88, 0.21}
\definecolor{canaryyellow}{rgb}{1.0, 0.94, 0.0}
\definecolor{daffodil}{rgb}{1.0, 1.0, 0.19}
\definecolor{electricyellow}{rgb}{1.0, 1.0, 0.0}

\lstdefinestyle{myJava}{ 
  frame=tb,
  language=Java,
  showstringspaces=false,
  columns=flexible,
  basicstyle={\small\ttfamily},
  numbers=none,
  numberstyle=\tiny\color{gray},
  keywordstyle=\color{blue},
  commentstyle=\color{dkgreen},
  stringstyle=\color{mauve},
  breaklines=true,
  breakatwhitespace=true,
  tabsize=3,
  morekeywords={Mono,Optional,Tuple4,Future,List,JsonObject,Map}, 
}

\definecolor{bluekeywords}{rgb}{0.13,0.13,1}
\definecolor{greencomments}{rgb}{0,0.5,0}
\definecolor{redstrings}{rgb}{0.9,0,0}

\lstdefinestyle{mySharpC}{
  frame=tb,
  language=[Sharp]C,
  showstringspaces=false,
  columns=flexible,
  basicstyle={\small\ttfamily},
  keywordstyle=\color{bluekeywords},
  commentstyle=\color{greencomments},
  stringstyle=\color{redstrings},
  breaklines=true,
  breakatwhitespace=true,
  tabsize=3,
  showspaces=false,
  showtabs=false,
  morekeywords={async,await,var},    
}

\lstdefinelanguage{rock}{
keywords={SUM,IFF,GROUP,BY},
morekeywords={for,each,FOR,EACH,WHERE,AFTER,PRECEDES},
keywordstyle=\color{blue},
moredelim=[is][\textbf]{|}{|},
basicstyle=\small\ttfamily\linespread{0.1},
frame=tb,
escapeinside={(*}{*)},
columns=fullflexible,
}

\usepackage{subfig}

\usepackage{enumitem}

\usepackage{csquotes}

\usepackage{comment}

\usepackage{tikz,pgfplots}
\pgfplotsset{compat=1.7} 

\usepackage{tabularx}
\usepackage{boxedminipage}
\newcolumntype{Y}{>{\centering\arraybackslash}X}
\newcolumntype{M}{>{\centering\arraybackslash}m}

\if 1\preprint
\else
\newcommand\vldbdoi{XX.XX/XXX.XX}
\newcommand\vldbpages{XXX-XXX}
\newcommand\vldbvolume{14}
\newcommand\vldbissue{1}
\newcommand\vldbyear{2020}
\newcommand\vldbauthors{\authors}
\newcommand\vldbtitle{\shorttitle}
\newcommand\vldbavailabilityurl{}
\fi

\begin{document}


\if 1\preprint
\title{Data Management in Microservices: State of the Practice, Challenges, and Research Directions (Extended Version)}
\else
\title{Data Management in Microservices: State of the Practice, Challenges, and Research Directions}
\fi



\author{Rodrigo Laigner}
\affiliation{%
  \institution{University of Copenhagen}
  \city{Copenhagen}
  \state{Denmark}
}
\email{rnl@di.ku.dk}

\author{Yongluan Zhou}
\affiliation{%
  \institution{University of Copenhagen}
  \city{Copenhagen}
  \state{Denmark}
}
\email{zhou@di.ku.dk}

\author{Marcos Antonio Vaz Salles}
\affiliation{%
  \institution{University of Copenhagen}
  \city{Copenhagen}
  \state{Denmark}
}
\email{vmarcos@di.ku.dk}

\author{Yijian Liu}
\affiliation{%
  \institution{University of Copenhagen}
  \city{Copenhagen}
  \state{Denmark}
}
\email{liu@di.ku.dk}

\author{Marcos Kalinowski}
\affiliation{%
  \institution{PUC-Rio}
  \city{Rio de Janeiro}
  \country{Brazil}
}
\email{kalinowski@inf.puc-rio.br}

\begin{abstract}


Microservices have become a popular architectural style for data-driven applications, given their ability to functionally decompose an application into small and autonomous services to achieve scalability, strong isolation, and specialization of database systems to the workloads and data formats of each service.
Despite the accelerating industrial adoption of this architectural style, an investigation of the state of the practice and challenges practitioners face regarding data management in microservices is lacking. To bridge this gap, we conducted a systematic literature review of representative articles reporting the adoption of microservices, we analyzed a set of popular open-source microservice applications, and we conducted an online survey to cross-validate the findings of the previous steps with the perceptions and experiences of over 120 experienced practitioners and researchers.

Through this process, we were able to categorize the state of practice of data management in microservices and observe several foundational challenges that cannot be solved by software engineering practices alone, but rather require system-level support to alleviate the burden imposed on practitioners. 
We discuss the shortcomings of state-of-the-art database systems regarding microservices and we conclude by devising a set of features for microservice-oriented database systems.

\end{abstract}




\maketitle

\if 1\preprint
\else
\begingroup\small\noindent\raggedright\textbf{PVLDB Reference Format:}\\
\vldbauthors. \vldbtitle. PVLDB, \vldbvolume(\vldbissue): \vldbpages, \vldbyear.\\
\href{https://doi.org/\vldbdoi}{doi:\vldbdoi}
\endgroup
\begingroup
\renewcommand\thefootnote{}\footnote{\noindent
This work is licensed under the Creative Commons BY-NC-ND 4.0 International License. Visit \url{https://creativecommons.org/licenses/by-nc-nd/4.0/} to view a copy of this license. For any use beyond those covered by this license, obtain permission by emailing \href{mailto:info@vldb.org}{info@vldb.org}. Copyright is held by the owner/author(s). Publication rights licensed to the VLDB Endowment. \\
\raggedright Proceedings of the VLDB Endowment, Vol. \vldbvolume, No. \vldbissue\ %
ISSN 2150-8097. \\
\href{https://doi.org/\vldbdoi}{doi:\vldbdoi} \\
}\addtocounter{footnote}{-1}\endgroup

\ifdefempty{\vldbavailabilityurl}{}{
\vspace{.3cm}
\begingroup\small\noindent\raggedright\textbf{PVLDB Availability Tag:}\\
All relevant code, and synthetically created data, will be made publicly available on an institutional repository upon acceptance. 
\endgroup
}
\fi

\section{Introduction}
\label{ref:intro}

\begin{figure*}
\centering
 \subfloat[\centering Traditional monolithic architecture]{{
  \includegraphics[width=0.41\textwidth]{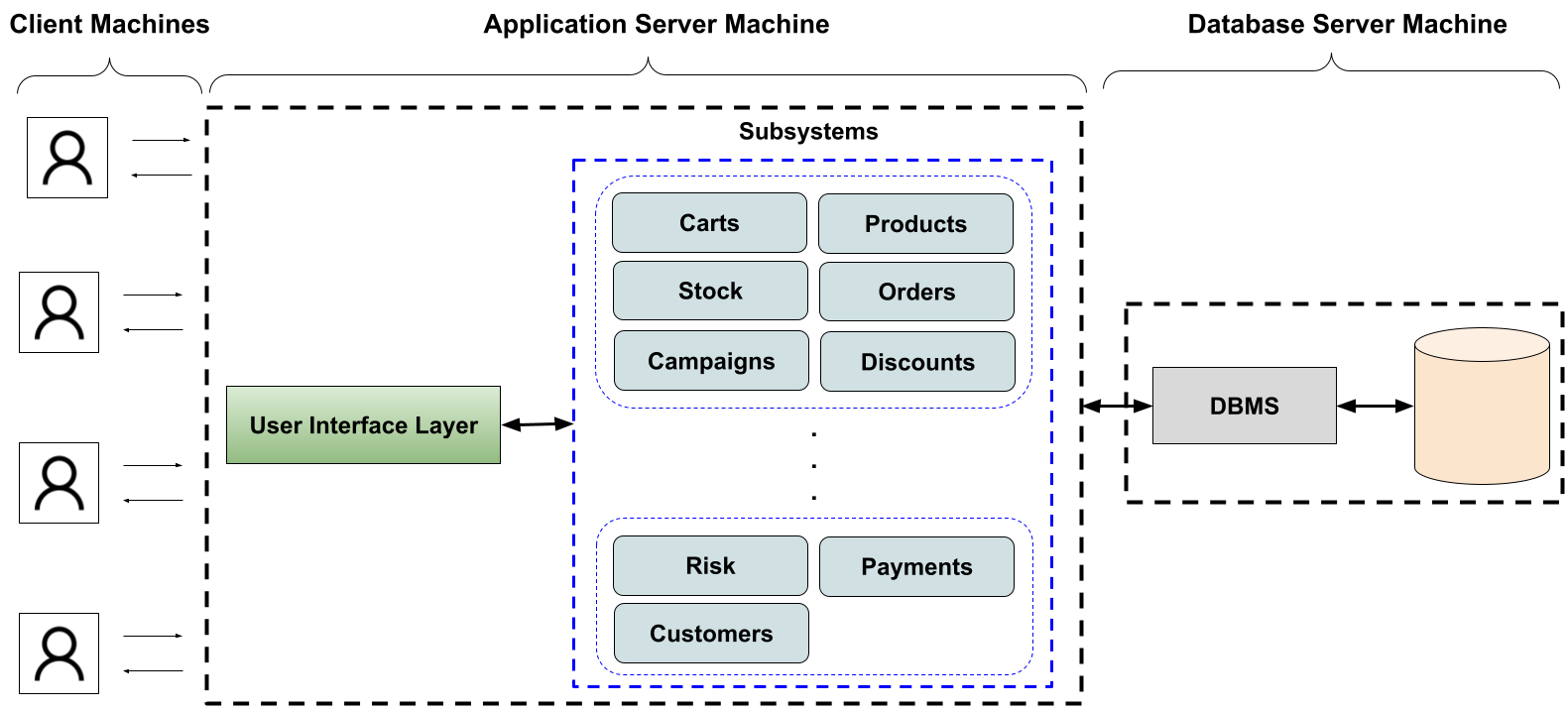}
  }}
  \qquad
\subfloat[\centering Microservice architecture]{{
  \includegraphics[width=0.535\textwidth]{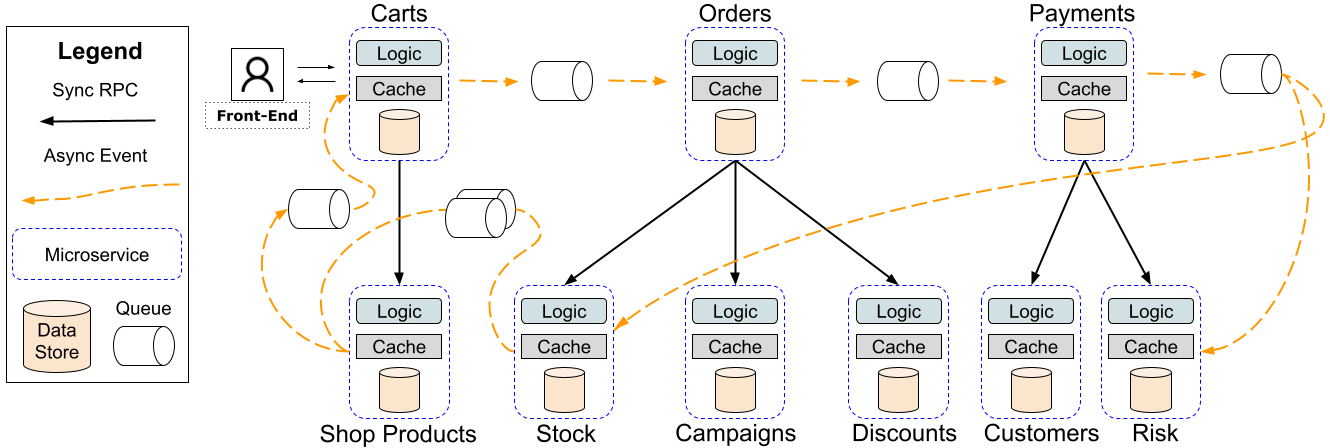}
  }}
  \vspace{-1ex}
\caption{Monolithic vs. microservice architectures}
\label{fig:diff_mono_msa}
\vspace{-2ex}
\end{figure*}


The advent of large-scale online services provoked an architectural shift in the design of data-driven applications, with resulting needs for designing distributed application systems from the point of views of both computational resources and software development team organization \cite{fowler:14,newman:15}. In particular, we are witnessing the increasing adoption of microservice architectures to replace the traditional monolithic architecture (Figure~\ref{fig:diff_mono_msa}). In contrast to a monolithic architecture (Figure~\ref{fig:diff_mono_msa}(a)), where modules and/or subsystems are integrated and cooperate in a centralized manner, a microservice architecture, (Figure~\ref{fig:diff_mono_msa}(b)), organizes an application as a set of small services that are built, deployed, and scaled independently. 

In the scenario of Figure~\ref{fig:diff_mono_msa}, we depict an e-commerce application to contrast both designs. In the monolithic case, to process an order, for example, the \textit{Cart} module performs a function call to the \textit{Order} module, which then performs additional function calls to the \textit{Stock}, \textit{Campaigns}, and \textit{Discounts} modules to safeguard the order is correctly placed. In contrast to direct function calls between modules, microservices communicate with each other through remote calls, such as HTTP-based protocols~\cite{fowler:14,gan2019open} or asynchronous messages~\cite{zimmermann:17}. In a microservice-based application, a new functionality or bug fix does not require a new build and subsequent deployment of the whole application, but instead, it can be managed by redeployment of a microservice unit. In the same line of reasoning, microservices also enable design for failure. As faults in individual microservices are isolated, their propagation to other building blocks of the architecture is limited~\cite{zimmermann:17}.

Furthermore, each microservice may also manage its own database that is suitable to the data formats and workloads of the microservice. This flexibility is often associated with the polyglot persistence principle, where different categories of database systems (e.g., loosely structured NoSQL vs. relational) service separate microservices \cite{fowler_polyglot_persistence}. For instance, the \textit{Products} database may take advantage of a document-oriented model to allow for agile schema evolution, whereas the \textit{Stock} microservice may rely on the relational model to safeguard constraints over stock items. 
As a result, a microservice architecture represents a significant shift from traditional monolithic transaction processing systems. In particular, in the monolithic architectural style, transactions can be easily executed across modules, while, in microservices, it becomes necessary to break these transactions down due to the decomposition of the application into small parts.

\noindent\textbf{Motivation.} Despite the increased adoption of microservices in industry settings \cite{8712154,7300793,HasselbringS17,gouigoux:2017,8004304,esposte:2017,mazzara:2017,azevedo:2019,8712376} and the perception that data management is a major challenge in microservices \cite{Holger:19,viggiato:2018,francesco:2017,luz:18,8703917,soldani2018pains}, there is little research on the characteristics of data management in microservices in practice. Besides, existing studies provide a limited investigation of the major challenges practitioners face regarding data management in such an architectural style. For instance, it is unclear which database technologies and patterns are adopted, which data consistency semantics are employed within and across services, or which mechanisms are used to exchange data in these architectures. Understanding these issues would provide valuable insights as to how to advance data management technologies to meet the needs of microservice applications.



\noindent\textbf{Methodology.} To bridge this gap, this paper presents an investigation of the state of the practice of data management in microservices. Specifically, we perform an exploratory study based on the following methodology: (i) we systematically review the literature on articles reporting the adoption of microservice architectures. From 300 peer-reviewed articles analyzed, 10 representative articles~\cite{mazzara:2017,esposte:2017,azevedo:2019,8004304,HasselbringS17,7300793,8712154,ciavotta2017microservice,Viennot:15,LaignerKLSO20} were selected for review; (ii) we analyze 9 popular microservice-based applications~\cite{ftgo,sentilo,LakesideMutual,eShopOnContainers,vertx,pitstop,petclinic,event_stream,sockshop}, selected out of more than 20 open-source projects, and; (iii) we design an online survey to gather the opinions of developers and researchers experienced with microservices in real-world settings, allowing us to cross-validate the findings of the previous steps. In total, more than 120 practitioners provided important information about their microservices' deployments in industry settings. Taken together, these three interrelated explorations provide new and comprehensive evidence on data management practices and challenges in microservices. Additional details about our methodology can be found in~\cite{extended}.


\noindent\textbf{Findings.} From our investigation, we observed that microservice developers are dealing with a plethora of data management challenges. While microservices are supposed to work autonomously, they often surprisingly end up exhibiting functionality and private state dependencies amongst each other. 



As a result, developers have a hard time reasoning about enforcement of application safety within and across microservices. The latter relates to challenges in managing constraints over distributed microservice states, reasoning about the unintended interleaving of event streams, dealing with weak concurrency isolation, enforcing data replication semantics, and ensuring consistency across a variety of storage technologies. 

Practitioners are poorly served by state-of-the-art database systems (DBMSs) and end up weaving together several heterogeneous data systems such as message brokers, in-memory caches, analytical engines, and loosely structured and structured DBMSs in an ad-hoc manner. This system complexity leads to a substantial amount of data management logic at the application layer to meet the data management requirements of microservices.


As a consequence, database systems no longer play a central role in this novel paradigm, often relegated to providing data storage functionality. The lack of a holistic view, by means of an unawareness of the dataflow, constraints, and the complex interplay among microservices, leads to the impossibility of effectively ensuring data and application safety in the microservice paradigm.








\noindent\textbf{Contributions.} In summary, we make the following contributions:

\noindent(i) We survey a gamut of experienced practitioners, the literature, and popular open-source microservice applications to characterize the state of the practice of data management in microservices. The findings (presented along $\S$~\ref{sec:state_of_practice}, $\S$~\ref{sec:computations}, $\S$~\ref{sec:microservices_glass}, and $\S$~\ref{sec:pains}) reveal several ad-hoc data management practices in microservice architectures never reported before in the literature, suggesting that developers are insufficiently served by state-of-the-art DBMSs.
\newline
\noindent(ii) We illustrate through source code snippets, extracted from popular open-source microservices, challenges that practitioners face when trying to meet the data management requirements of microservices ($\S$~\ref{sec:microservices_glass}). These challenges are also highlighted and extended by experienced microservice developers when asked about their most pressing pains regarding data management ($\S$~\ref{sec:pains}), thus illustrating issues that database technology should aim to address.
\newline 
\noindent(iii) Based on the observed data management practice and challenges, we present a set of features for future database systems to tackle the data management requirements of microservices by design, so they can play a central role in this new paradigm ($\S$~\ref{sec:towards}). We discuss why the state of the art is not able to address all the challenges in conjunction and we conclude by suggesting how the features can be realized into a microservice-oriented DBMS. 


The three contributions together provide new directions for the database community to start leading the efforts on data management for microservices. Our goal is to inform engineers of future data management systems about the unmet needs of an emerging application paradigm by providing comprehensive evidence of shortcomings and pitfalls microservice developers face when dealing with data management. We view this work as an example of engaging with database system users to understand and characterize the practice in order to derive new research opportunities for emerging applications. We hope the results drive the reflection of our community towards effectively meeting the needs of microservice-based applications.


\section{Related Work}
\label{sec:related_work}

Despite the extensive work in microservice architectures in other communities, an in-depth analysis of data management in this context is lacking in existing research. Works in software engineering focus on migrating from monolithic architecture to microservices \cite{carvalho2019analysis,carrasco:18} or investigating other general software engineering aspects \cite{ghofrani2018challenges,viggiato:2018,francesco:2017,8703917,Holger:19}, such as software attributes (e.g., coupling and cohesion). To the best of our knowledge, our work is a first step towards understanding how microservice developers interact with the database systems that the data management community builds, fostering further research. 



In addition, our work provides an in-depth characterization of challenges faced by microservice developers while implementing data management logic in the application-tier. In regard to this contribution, although some studies described related pitfalls, such as \textit{shared persistence}~\cite{8354414,IlariaSmells2020}, and previous literature investigated architectural smells and anti-patterns in microservices~\cite{neri:19,carrasco:18,taibi:2019,microservicesAntipatterns}, they fail to capture properties of consistency models and technical issues of database systems, such as data replication and constraint enforcement, as we provide in this paper. Most importantly, in light of the many unveiled data management challenges in microservice architectures, our work is the first to reflect on core limitations that prevent database systems from adequately serving the needs of this emerging architectural style.
\section{State of the Practice}
\label{sec:state_of_practice}


In this section, we focus on characterizing the motivating factors for data management in microservices as well as the most prominent DBMSs and deployment patterns employed.


\subsection{Motivating factors}
\label{subsec:motivations}

\begin{table}
\centering
\caption{Motivations for microservices in data management}
\vspace{-3ex}
\begin{tabularx}{\columnwidth}{|Y|c|c|}
\hline
\textbf{Motivation} & \textbf{\#} & \textbf{\%} \\
\hline
Scalability through functional decomposition & 55 & 35.71 \\
\hline
Fault-isolation (e.g., increasing data availability) & 32 & 20.77 \\
\hline
Agility on data change (e.g., facilitating schema evolution) & 32 & 20.77 \\
\hline
To enable event-driven data management (e.g., as opposed to classic pull-based data querying) & 23 & 14.93 \\
\hline
Polyglot persistence & 9 & 5.84 \\
\hline
Others & 3 & 1.94 \\
\hline
\end{tabularx}
\label{tab:motivations}
\vspace{-3ex}
\end{table}

Existing works \cite{fowler:14,gan2019open,carvalho2019analysis,HasselbringS17,newman:15} argue that the major reasons for adopting microservices are related to fault-isolation, independent software evolution (including schema evolution), and scalability of individual system components. By investigating several microservice deployments, we were able to reveal that such desirable properties are enabled by data partitioning and decentralized data management -- particularly by means of use of a database/schema per microservice. Thus, these motivations are intrinsically related to data management, and decentralized data management is a major foundation in the adoption of microservices.



To investigate the most compelling directions for future avenues of research in data management for microservices, we asked the survey participants to select the top 2 reasons to adopt a microservices architecture regarding data management. The 5 options given were centered on data management concerns (i.e., no software engineering concerns, such as loose coupling and easier maintenance, were considered). The provided options were influenced by the following: (i) a preliminary assessment with four industry representatives with strong background in microservices through interviews; (ii) the literature review; (iii) the analysis of open-source repositories; and (iv) discussions among the authors. Table~\ref{tab:motivations} shows the options provided and respective responses. We highlight the following important observations (referenced hereafter by \textbf{O\#}).





\noindent\textbf{Functional partitioning:} To support scalability (i.e., spreading functional groups across databases) 
and high data availability (i.e., achieving functional isolation of errors), functional decomposition of the application is a major driver for adopting microservices according to both survey respondents (57\%) and the literature~\cite{mazzara:2017,esposte:2017,8004304,HasselbringS17,ciavotta2017microservice,8712154,LaignerKLSO20}. 
Though not explicitly mentioned by literature, we deduce from our analysis that functional decomposition is reminiscent of the idea of functional scaling introduced by Pritchett \cite{pritchett:08}, a strategy that involves "grouping data by function and spreading functional groups across [different] databases."

\vspace{1ex}
\noindent\fbox{\begin{minipage}{26em}
\textbf{O1.} The results suggest that the design of data management technologies for microservices should focus not only on scalability, but also on stronger mechanisms for fault-tolerance and error isolation.
\end{minipage}}
\vspace{1ex}

\noindent\textbf{Decentralized data management:} Practitioners developing applications in real-world settings largely deal with evolving requirements and subsequent schema changes \cite{Viennot:15}. The ability of microservice architectures to provide independently-evolving schemas in different services, in contrast with the unified schema of monolithic architectures, is another major driver~\cite{esposte:2017,HasselbringS17,7300793,ciavotta2017microservice,Viennot:15}. Surprisingly, although polyglot persistence derives naturally from decentralized data management, it is the least cited motivation in both literature \cite{azevedo:2019,7300793,Viennot:15} and the survey, being roughly 3.5 times less cited by practitioners than schema evolution.

\vspace{1ex}
\noindent\fbox{\begin{minipage}{26em}
\textbf{O2.} The results suggest that the support for a large variety of data models is less of a pressing need than schema evolution in microservices.
\end{minipage}}
\vspace{1ex}

\noindent\textbf{Event-driven microservices:} Event-driven systems constitute an emerging trend in the design of data-driven software applications~\cite{kleppmann:2019}. Microservices are often mentioned as a compelling paradigm for designing event-driven applications~\cite{katsifodimos2019operational,vivekshahthesis,LaignerKLSO20}. Almost 15\% of the participants consider event-driven data management a primary motivation for microservices, which may indicate a trend in industry adoption. Most papers~\cite{mazzara:2017,esposte:2017,8004304,HasselbringS17,7300793,ciavotta2017microservice,Viennot:15,LaignerKLSO20} mention the use of asynchronous primitives for message-oriented communication to achieve loose coupling among microservices and facilitate decentralized data management. The same trend is encountered in open-source repositories~\cite{vertx,sentilo,event_stream,ftgo,eShopOnContainers,ftgo,LakesideMutual}.



\vspace{1ex}
\noindent\fbox{\begin{minipage}{26em}
\textbf{O3.} The results suggest that asynchronous events should emerge as a concern when devising data management technologies for microservices.
\end{minipage}}
\vspace{1ex}


\noindent\textbf{Summary.} Our results indicate that functional decomposition, fault isolation, schema evolution, and event-driven architecture are the primary reasons behind the adoption of microservices for data management. 

\subsection{Database systems and deployment patterns}
\label{subsec:database_patterns}

There are three mainstream approaches for using database systems in microservice architectures: (i) private tables per microservice, sharing a database server and schema; (ii) schema per microservice, sharing a common database server; and (iii) database server per microservice \cite{messina:2016}. We asked the participants which database patterns they use to support data management in their microservices. We also asked the participants what drivers led to the adoption of each database pattern in their microservice architectures. 

We observe that most practitioners prefer encapsulating a microservice's state within its own managed database server (43\% of responses).
The same trend is observed in the literature and open-source repositories. The participants indicated the following drivers that lead to this adoption: 
(i) achieving loosely-coupled microservices; (ii) independent data layer scalability, which would otherwise be challenging with a single database server supporting multiple (heterogeneous) tenant applications; and (iii) fault-isolation, which is naturally derived from the already mentioned formation of independent silos of data.

Besides, practitioners, literature, and open-source repository analysis indicate that container-based deployment is the de-facto practice. Each microservice and respective database are bundled in separate containers, thus guaranteeing that each can be scaled independently and faults are limited to the container boundary. 

\vspace{1ex}
\noindent\fbox{\begin{minipage}{26em}
\textbf{O4.} The results suggest that microservices are prevalently deployed in individual containers, predominantly using the database-per-microservice pattern to achieve performance and fault isolation.
\end{minipage}}
\vspace{1ex}




Furthermore, we asked the participants what database systems they have been adopting in their microservice-based applications. Our objective was to understand the types of database technologies adopted, specially concerning the data model, performance, and scalability aspects. 
The adoption of multiple DBMS belonging to at least two of the provided categories is often reported. For instance, 47.97\% indicated the use of at least one relational DBMS in conjunction with MongoDB. Besides, we observed a trend (15\%) of use of the following stack: a relational DBMS (e.g., PostgreSQL, MySQL, or SQL Server) + Redis + MongoDB + ElasticSearch. 

The most common use case for this stack is the use of relational or document-oriented DBMSs for the underlying microservice databases, Redis as a caching layer to provide fast data access to recurring requests, and replication of data through an event-driven approach to ElasticSearch for fast online analytical queries. Overall, the results of DBMS adoption from the survey are very aligned with the reviewed papers and the open-source~repositories.

\section{Data Management Logic In Practice}
\label{sec:computations}

\begin{figure}[t!]
    \centering
    \begin{tikzpicture}\begin{axis}
            [
            xbar, xmin=0,xmax=100,width=8cm,
            height=3.5cm,
            xlabel={\% of participants},
            y tick label style={
                font=\small,
                text width=1.25cm,
                align=center,
            },
            symbolic y coords={BT,ED,OQ,SP},
        ]
        \addplot coordinates{(95.84,BT) (66.67,ED) (39.39,OQ) (25.76,SP) };
        \end{axis}
    \end{tikzpicture}
    \vspace{-3ex}
    \caption{Data management tasks indicated by participants}
    \label{fig:computations}
    \vspace{-3ex}
\end{figure}
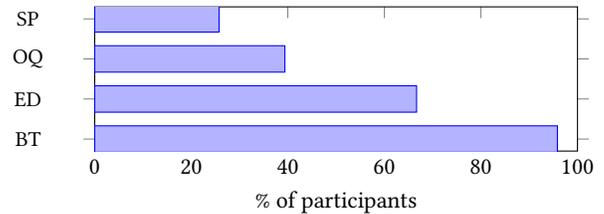

Continuing from our last section, we were also interested in characterizing the types of queries and data management logic that are performed in microservice-based applications. 
Thus, to avoid the threat of misconceptions in answers, we defined a set of open and multiple-choice questions to obtain from the participants relevant information about 
how business logic is implemented in their applications. Figure~\ref{fig:computations} shows the types of data management logic identified from the responses, characterized by Cross-Microservice Business Transactions (BT), Event-Driven Computing (ED), Online Queries (OQ), and Stream Processing (SP).  


\subsection{Cross-microservice business transactions}\label{subsec:business_transactions}

The studies reviewed from literature \cite{mazzara:2017,Viennot:15,8004304,HasselbringS17,8712154}, the open-source repositories \cite{eShopOnContainers,event_stream,LakesideMutual,vertx,pitstop,petclinic,sockshop,ftgo}, and the use cases described by participants indicate the prevalence of decentralized OLTP-like workloads in microservice-based applications, such as tracking orders in web shop applications.

\noindent\textbf{Evidence from literature and open-source repositories.} By analyzing the open-source repositories, we observed that OLTP-like intra-microservice transactions are common, such as those found in conventional applications \cite{chen2014detecting,bailis:15}. Similar to Pavlo's findings \cite{pavlo2017we}, we did not find evidence of the use of stored procedures or the configuration of databases in serializable isolation. The same applies to literature. 


Regarding business transactions across microservices,~\footnote{Our use of the term "business transaction" in this paper refers to any unit of work carried out across microservices~\cite{ONeil18}, not necessarily under ACID guarantees.} conversations (i.e., the interaction between a set of consumer and producer services) are prevalent. Hohpe~\cite{hohpe2007let} argues that orchestration and choreography are the two types of interactions that take place in the context of distributed web applications. Most papers~\cite{7300793,HasselbringS17,8004304,azevedo:2019,esposte:2017,mazzara:2017,LaignerKLSO20} and open-source projects~\cite{LakesideMutual,eShopOnContainers,sentilo,vertx,petclinic,sockshop,pitstop,event_stream} report the use of the choreography conversation pattern \cite{hohpe2007let} through both synchronous and asynchronous event-based workflows.

In open-source repositories, event-based workflows are dominant, implying that updates and operations affecting other microservices are queued for asynchronous processing. This finding has led us to observe that microservice architectures indeed follow the BASE model~\cite{pritchett:08}, which targets functionally decomposing an application to achieve higher scalability in exchange for a weak consistency model. The same trend is found in the literature. Besides, some papers report the use of orchestration~\cite{8712154,ciavotta2017microservice,azevedo:2019}. Only one open-source project~\cite{ftgo} adopts a saga-like orchestration. 
In any case, the options above are characterized by weak consistency models~\cite{bailis13} where on an operation that spans multiple microservices, the tasks may complete at any point in time, and the data returned is only eventually consistent.

While the literature~\cite{fowler:14,zimmermann:17,newman:15} mentions the principle that microservices are autonomous components that are independently deployed and evolved, we observed that most microservice-based applications often perform operations that span multiple microservices (Figure~\ref{fig:computations} and Table~\ref{tab:mechanisms}), which indicates a functionality dependence between microservices.
However, consistent with recent discussions in the community~\cite{pavlo2017we,helland_life_beyond,bailis_interview,data_outside}, we did not find any evidence of distributed transactions, such as through the 2-Phase Commit (2PC) protocol, in both the open-source repositories and the reviewed papers from literature. 


\vspace{1ex}
\noindent\fbox{\begin{minipage}{26em}
\textbf{O5.} Distributed commit protocols do not enjoy popularity in microservices. 
The lack of effective and intuitive application abstractions may play a role in refraining practitioners from adopting such protocols~\cite{helland_life_beyond,data_outside}.
\end{minipage}}
\vspace{1ex}


\noindent\textbf{Evidence from industry settings.} To further characterize the implementation of business transactions in microservices, we focused on understanding how consistency guarantees are enforced in industrial settings. We asked the participants which mechanisms to coordinate operations spanning multiple microservices they have been employing. The options were defined based on the patterns found in the literature, open-source repositories, along with a preliminary assessment of the survey with industry representatives.\footnote{By investigating the literature, we have found that the patterns for inter-microservice operations are not homogeneously defined. Besides, the preliminary assessment confirmed that developers tend to mark this option in a specialized way, e.g., centralized sagas are infrequently identified as orchestration or decentralized sagas as choreography. We then opted to proceed with different options, even though some may be subclasses of others.} Table~\ref{tab:mechanisms} shows the responses sorted in descending order.




%

\begin{table}
\centering
\caption{Mechanisms for inter-microservice coordination}
\vspace{-3ex}
\begin{tabularx}{\linewidth}{|Y|c|c|}
\hline
\textbf{Coordination mechanism} & \textbf{\#} & \textbf{\%} \\
\hline
Orchestration & 37 & 22.84 \\
\hline
Sagas (centralized approach, with a Saga coordinator) & 24 & 14.81 \\
\hline
The Back-end for Front-end Pattern (BFF) & 24 & 14.81 \\
\hline
Choreography & 22 & 13.58 \\
\hline
Sagas (decentralized approach, i.e., no Saga coordinator) & 14 & 8.64 \\
\hline
Distributed transactions (e.g., via 2PC) \footnotemark & 14 & 8.64 \\
\hline
2-Phase Commit & 11 & 6.79 \\
\hline
Others & 16 & 9.88 \\
\hline
\end{tabularx}
\label{tab:mechanisms}
\vspace{-3.5ex}
\end{table}

\footnotetext{Some respondents were selecting the option \textit{distributed transactions} but their subsequent answers were not compatible with distributed commit protocols (e.g., employing events to trigger actions asynchronously as state becomes consistent in a microservice). We then opted to change the option to \textit{2-Phase Commit}. By analyzing such answers, we estimate that 11 of the responses are not compatible with distributed transactions. This would result in 33 responses (20.37\%) being attributable to choreography in total.}

In contrast with findings from literature and open-source repositories, orchestration-like (including sagas \cite{molina:87} and the backed-end for front-end pattern (BFF)\footnote{In this mechanism, a service centralizes the role of performing requests across microservices.}~\cite{bff,bff2,bff3}) mechanisms are the most popular in industry settings. To further understand these orchestration-like mechanisms, we asked the participants which orchestration engines they used to support operations spanning multiple microservices. The results highlight that the adoption of custom-made (e.g., company-built) orchestration engines is prevalent among participants (51.2\%).


We also asked the participants to briefly describe one of their use cases involving consistency in operations spanning multiple microservices. 32 out of 90 (35.5\%) responded and most responses (78\%) indicated the implementation of workflows through application code and the use of application-level validations to safeguard the constraints of the workflow.

Despite the prevalence of orchestration-like mechanisms (22.8\%), choreographies are also highly mentioned (20.4\%, see footnote~3). An interesting quote provided by one of the respondents characterizes how choreographies are implemented in industry, which aligns well with our findings in literature and open-source applications:
\vspace{-0.1ex}
\begin{displayquote}
I am absolutely against the business logic inside the database. Depending on the scale I would refrain from using transactions at all, favoring an event-driven approach, with eventual consistency and micro-transactions.
\end{displayquote}
\vspace{-0.1ex}

The main difference between the orchestration-like mechanisms described by the participants and the choreography mechanisms found in the open-source repositories, literature, and participants is the type of communication. The former is synchronous and HTTP-based, whereas the latter is mostly event-based and asynchronous.


A substantial fraction of the surveyed papers~\cite{mazzara:2017,esposte:2017,8004304,HasselbringS17,8712154,ciavotta2017microservice,Viennot:15} employ choreography as the mechanism for cross-microservice coordination and do so by means of the implementation of asynchronous and event-based workflows. This pattern is consistent with the responses obtained from practitioners, as choreography is highlighted as one of the most prevalent mechanisms (Table~\ref{tab:mechanisms}). Most analyzed open-source repositories~\cite{eShopOnContainers,ftgo,sentilo,vertx,pitstop,sockshop,event_stream,LakesideMutual} also show the same pattern.




\vspace{1ex}
\noindent\fbox{\begin{minipage}{26em}
\textbf{O6.} The results suggest the prominence of orchestration-like mechanisms in industry settings, in contrast to the prevalence of choreography in the open-source repositories. Additionally, 2PC is not used often, while asynchronous and event-based coordination is the norm.
\end{minipage}}


\subsection{Online queries} \label{subsubsec:online_querying}
While one may argue that queries spanning multiple microservices are antagonistic to the principles of state encapsulation and independent data silos of microservices, we found abundant evidence of such cases in the literature~\cite{azevedo:2019,LaignerKLSO20,ciavotta2017microservice,Viennot:15} and open-source repositories~\cite{ftgo,eShopOnContainers,petclinic,sentilo,vertx}. Therefore, to further understand this trend, we asked the survey participants to identify if they have implemented queries aggregating data from multiple microservices and to describe one of their use cases. 27 (40.3\%) of the participants declared the use of some mechanism and 22 of them (81.4\%) provided a short-answer describing it. We unveiled three mechanisms to allow for such queries and explain them as follows.

\noindent \textbf{A. Queries aggregating data belonging to different microservices.} In this mechanism, a consumer service contacts, often through HTTP requests, a set of microservices through their APIs. After receiving all responses, the consumer service then aggregates the data in-memory (also performing joins, if necessary) and serves the client. We identified the following three practices to implement such a mechanism: (i) Six respondents described the use of composition of service calls, i.e., a microservice performs the necessary synchronous requests to retrieve data from other microservices. 
(ii) One respondent declared the use of the BFF pattern~\cite{bff}. We also identified such practice in a repository~\cite{eShopOnContainers}. 
(iii) Lastly, one respondent declared the use of the API Gateway pattern~\cite{api_gateway:20}. We also observed its adoption in open-source repositories~\cite{ftgo,petclinic,vertx} and the literature~\cite{azevedo:2019,7300793}. 
From the respondents' answers, we could not observe significant differences between the BFF and the API Gateway patterns in terms of query serving. 

\noindent\textbf{B. Replication.} We also unveiled the use of \textit{ad-hoc} mechanisms for data replication across microservices for online querying purposes. We explain the identified practices as follows.

(a) \textit{Replication across microservices.} This practice is characterized by a microservice generating events related to its own updated data items and communicating these changes asynchronously, often through persistent messaging supported by a message broker. In 7 out of 9 (77 \%) projects analyzed~\cite{vertx,sentilo,event_stream,ftgo,eShopOnContainers,ftgo,LakesideMutual}, we found code fragments used for communication-based replication that presuppose weak delivery semantics. In other words, although updates to the same object are often sequentially ordered by the publisher, there is no ordering guarantee regarding updates to different replicated objects. As a result, causal dependencies are ignored on updating replicated data items. The responses provided by 5 participants in open answers suggest the same trend. This choice appears to be consistent with the eventual consistency semantics adopted by the synchronous query mechanisms described above.


(b) \textit{Replication to a database.} This practice is characterized by two mechanisms:
(i) Daemon workers, one for each microservice and its respective generated events, or a central service, are responsible for subscribing to data item updates and replicating these to a special-purpose database used for querying; (ii) this practice is also characterized by the use of batch workers (usually special-purpose microservices) to extract data from microservices periodically (with a pull-oriented strategy) and replicate those in a neutral data repository for fast querying (e.g., ElasticSearch). We suspect the second approach is reminiscent of the behavior of Extraction-Transform-Load (ETL) tools~\cite{vassiliadis2009survey}. Although it is unknown why ETL tools are not being employed for such task, we believe the dynamicity provided by the autonomous deployment of microservices plays a role. In other words, microservices can be easily put to and removed from operation while not affecting others, whereas such a pattern is often not found in computational steps of ETL tools.

(c) Lastly, the use of data stream processing systems (DSPSs) for processing streams (e.g., updates to data items) generated by microservices to build materialized views was also mentioned. One respondent declared: ``[...] materializing views over various time windows, making them queryable to other services.'' 


\noindent\textbf{C. Views.} While service composition and replication are often subject to the adoption of the database per microservice pattern (Section \ref{subsec:database_patterns}), when microservices share the same database, practitioners may rely on views across multiple schemas to serve cross-microservice queries. This is the least cited practice.

\vspace{1ex}
\noindent\fbox{\begin{minipage}{26em}
\textbf{O7.} The results highlight that the decentralized data management principle does not refrain microservices from performing queries over distributed states. As a result, practitioners often rely on \textit{ad-hoc} mechanisms for data processing at the application-level.
\end{minipage}}

\subsection{Stream processing}
\label{subsubsec:stream_processing}

To understand how data stream processing systems (DSPSs) that the database community builds interact with microservice architectures, we asked the participants if they have already employed a DSPS in conjunction with microservices. We also asked them to provide a description of one of their use cases. 18 (26.86\%) out of 67 respondents declared the use of DSPSs and 13 (19.4\%) of them provided a description. We summarize these as follows.

\noindent\textbf{Data processing pipelines.} We observed the use of application libraries targeted at stream processing~\cite{kafka_streams} in microservices to perform data transformations, as mentioned by a participant: ``We use Kafka Streams to reorder (time window) out-of-sequence data[.]'' Besides, one respondent declared the formation of an ``ETL chain implemented with microservices that exchange information asynchronously using Kafka as a message bus. [...] We have several microservices in an ETL chain, [... including] several transformation steps.'' The person explains that ``the chain will fork at a certain point and one side of the fork will carry on the transformation up to message delivery to downstream systems, the other side of the fork will do asynchronous writes to an operational data store (a NoSQL database). [..] These writes are asynchronous to remove the DB from the critical path and ensure that messages can be delivered in near-realtime.''

These responses represent a surprising trend, since we did not find substantive evidence of the use of microservices for forming a data processing pipeline in the literature and open-source repositories. Existing literature suggests an impedance mismatch between microservices and DSPSs~\cite{katsifodimos2019operational,why_orleans_streams,wang2019modeling} particularly related to the stream processing abstraction. Often in the form of static dataflow graphs, operations such as filter, join, and aggregate are applied uniformly to all stream items in such abstraction. This model notably contrasts with microservice principles, including loose-coupling, fault-isolation, independent evolution, and autonomous deployment. 
Most importantly, the dynamic behavior of microservices, including operating over data items from different microservices, introduce a significant challenge to express complex business logic using this abstraction. 


\vspace{1ex}
\noindent\fbox{\begin{minipage}{26em}
\textbf{O8.} The responses suggest microservice developers find the static dataflow abstraction difficult to express their computations. As a result, they end up relying on the \textit{ad-hoc} formation of data processing pipelines by microservices.
\end{minipage}}
\vspace{1ex}






\noindent\textbf{Anomaly detection.} The use of DSPSs for anomaly detection based on event streams generated by microservices was mentioned by two respondents. For instance, one respondent declared the adoption of a stream processing engine for ``monitoring financial operations by analyzing situations and generating critical events for microservices that convert these events into alerts.'' 


\noindent\textbf{Replication and materialized views.} As already mentioned in the last section, four respondents indicated the use of stream processing engines to process data generated from microservices aiming at replicating data and materializing views.

\subsection{Event-driven computing}
\label{subsubsec:event_driven_computing}
As mentioned previously, we observed that microservice architectures often follow the BASE model~\cite{pritchett:08}, 
wherein organizing a computation in an event-driven architecture (EDA) is argued for scalability and architectural decoupling.
In an EDA, events that are relevant to an incoming request are generated when a consistent state is reached to allow further processing~\cite{pritchett:08}.
To characterize how EDA intersects with microservice architectures, we asked the participants to declare whether they have performed event-driven computations in microservices through an EDA and provide an example of one of their use cases in a short answer. A total of 45 (67.16\%) out of 67 participants indicated the use of such computations and 26 (38.81\%) provided a brief use case description. 

Overall, the responses are consistent with the findings from the literature and the open-source repositories. EDA is often an enabler of event-based and asynchronous workflows. Some quote snippets are provided as follows: (a) ``Pure choreography without central orchestrator;'' (b) ``natural way for microservices to collaborate when they depend on data from other microservices;'' (c) ``Payments system using an [EDA] to process ecommerce orders/payments.'' (d) ``[...] to trigger several microservices in our architecture.''


\section{Microservices through the looking glass}
\label{sec:microservices_glass}

In Sections~\ref{sec:state_of_practice} and \ref{sec:computations}, we observed that microservice applications deviate significantly from the architecture of traditional database applications, introducing significant data management logic at the application-level and a decentralized model for data management. 

Such findings urged us to investigate challenges faced by developers that would drive research avenues in data management. Particularly, we aim to answer whether the observed shift promotes challenges that cannot be solved by traditional database systems. 

Therefore, we present next a series of challenges (referenced by \textbf{C\#}) faced by developers that reveal a myriad of unmet needs that should be appropriately addressed by the database community.



\subsection{Cross-microservice validations}
\label{subsec:cross_validations}

\noindent \textbf{Background.} Given the ubiquity of business transactions across microservices, we start by reviewing a type of application-level validation employed by developers to ensure correctness in such cases. 
Take for example the snippet adapted from the project \textit{vertx} \cite{vertx} shown in Listing~\ref{lst:cross_microservice_validation}. Prior to proceeding with the client's checkout request, the \textit{cart} microservice verifies, through a HTTP remote call to the \textit{inventory} microservice whether the items in the cart are available. If so, it generates an event requesting the corresponding order to be processed.

However, such validation is not safe under concurrency. By the time the \textit{Order} microservice processes the event, one (or more) of the items in the cart might not be available anymore. Under high contention such criteria may lead to abusive generation of events, resource trashing, and may also introduce the burden to deal with compensation logic in the application. 
It is worthy to note we encountered such a pattern in several other projects, such as~\cite{ftgo,petclinic,LakesideMutual,eShopOnContainers,event_stream,pitstop}.

\lstset{style=myJava}
\begin{lstlisting}[caption={Cross-microservice validation example},label={lst:cross_microservice_validation},captionpos=b]
// Cart microservice: Checking availability of products
private boolean checkAvailableInventory(ShoppingCart cart) {
    List allInventories = getInventoryEndpoint()
            .HttpGet(cart.getProductItems());
    return allInventories.map(inventories -> {
          List insufficient = inventories.filter(item -> item.get("inventory") - item.get("amount") < 0).toList();
          if (insufficient.size() > 0) return false;
          else return true; });
}
\end{lstlisting}
\vspace{-1ex}


\noindent \textbf{Discussion.} In the case presented in Listing~\ref{lst:cross_microservice_validation}, coordination is a condition necessary to ensure correctness under conflicting reads and writes. However, in the absence of efficient solutions and intuitive interfaces for encoding 
concurrency control semantics
in the application-tier~\cite{data_outside}, developers end up encountering challenges to safeguard constraints across microservices.


Another impediment that makes the problem even more difficult comes from the heterogeneous database systems encountered in microservice architectures, the incompatible isolation levels, and the corresponding lack of interoperability among them.

\vspace{1ex}
\noindent\fbox{\begin{minipage}{26em}
\textbf{C1.} In the absence of efficient or viable solutions for isolation guarantees in the application-tier, microservice developers are exposed to concurrency anomalies. This creates a great barrier for expressing correctness criteria across different microservices.
\end{minipage}} 



\subsection{Implicit cross-microservice associations}
\label{subsec:implicit}




\lstset{style=myJava}
\begin{lstlisting}[caption={Implicit cross-microservice association example},label={lst:implicit},captionpos=b]
// Product microservice: manages available products
public void deleteProduct(String productId, Handler<AsyncResult<Void>> resultHandler) {
    this.removeOne(productId,DELETE_STATEMENT,resultHandler);
}
------------------------------------------------------------
// Cart microservice: product items without association
public class ShoppingCart {
  private List<ProductTuple> productItems;
  private Map<String, Integer> amountMap;
  public ShoppingCart() {}
} // additional code omitted
\end{lstlisting}
\vspace{-1ex}

\noindent \textbf{Background.} Even though microservices follow an approach similar to BASE for functional decomposition~\cite{pritchett:08}, developers do not refrain from modeling implicit associations across microservices.\footnote{We refer to "implicit associations" as those relationships between tables from different microservices that would exist if the application schema was designed as a single database.} We encountered several cases where enforcement of foreign key constraints across microservices is a necessary condition for ensuring correctness, and the applications analyzed show no evidence of such enforcement.

Consider the source code snippet exhibited in Listing~\ref{lst:implicit} adapted from \textit{event-stream}~\cite{event_stream}, \textit{vertx}~\cite{vertx}, and \textit{eShopContainers}~\cite{eShopOnContainers} applications. In the event of a product removal from the \textit{product} microservice, in the absence of cascading delete, operations carried out by the system over records stored in other microservices that rely on the existence of the deleted product(s) will still assume such a product exists, which may bring the system to an inconsistent state.
This pattern is observed across several other microservice applications~\cite{LakesideMutual,event_stream,eShopOnContainers,vertx,ftgo,pitstop}.


\noindent \textbf{Discussion.} Microservice developers have to explicitly enforce implicit foreign key constraints across microservices to avoid bringing the system into an inconsistent state, a complicated and error-prone task. However, in most cases, practitioners simply either ignore or are unaware of the consequences. Giving up foreign key constraint enforcement across microservices leads to "orphaned" records in one or more microservices. Such a pitfall is even worse than encoding associations under non-serializable isolation, since it may lead to a much higher number of anomalies~\cite{bailis:15}.

\vspace{1ex}
\noindent\fbox{\begin{minipage}{26em}
\textbf{C2.} The impossibility of declaring foreign key constraints between different microservices' schemas creates a great barrier for developers to enforce constraints across microservices.
\end{minipage}}
\vspace{-1ex}

\subsection{Cross-microservice queries}
\label{subsec:cross_queries}

\noindent \textbf{Background.} As part of our study, we observed the popularity of online queries in microservice architectures (Section~\ref{sec:computations}). Given the distributed nature of data stores, the data encapsulation may introduce challenges not found in traditional monolithic systems.

The literature \cite{azevedo:2019,7300793} and open-source repositories provide interesting examples of microservices being employed for data aggregation through queries spanning different microservices (and their underlying databases), often with different data models. Such implementations reveal a new trend on stateful middle-tier applications that are particular to microservices: encoding of data processing functionality at the application-level. 

Consider the example adapted from the project FTGO~\cite{ftgo}, shown in Listing~\ref{lst:query}. The code snippet exhibits a method (\textit{getOrderDetails}) responsible for reaching out to several microservices in order to consolidate in real-time a client view (i.e., the order details). 


\lstset{style=myJava}
\begin{lstlisting}[caption={Cross-microservice query example},label={lst:query},captionpos=b]
public OrderDetails getOrderDetails(Long orderId) {
    OrderInfo orderInfo = orderService.findOrderById(orderId);
    TicketInfo ticketInfo = kitchenService
            .findTicketById(orderId);
    DeliveryInfo deliveryInfo = deliveryService
            .findDeliveryByOrderId(orderId);
    BillInfo billInfo = accountingService
            .findBillByOrderId(orderId);
    Mono<OrderInfo,TicketInfo,DeliveryInfo,BillInfo> combined = Mono.zip(orderInfo, ticketInfo, deliveryInfo, billInfo);
    OrderDetails orderDetails = combined.map( OrderDetails::makeOrderDetails);
    return orderDetails;
}
\end{lstlisting}
\vspace{-2ex}


\noindent \textbf{Discussion.} Current state of the practice leaves the developer responsible for retrieving the appropriate data from each microservice and dealing with possible inconsistencies, such as fractured reads~\cite{bailis2016scalable}, which may lead to a complex code base and bugs. Besides, with such application-level data management there is no way to ensure that reads to different microservices reflect a view of the entire application at a single point in time. In other words, transactional consistency \cite{ports2010transactional} is not possible. Lastly, with such a sequential request pattern, as observed in Listing~\ref{lst:query}, some (or all) requests to microservices may fail, thus leading to missing records and an incomplete result. Such a pattern is also found in~\cite{eShopOnContainers,petclinic,LakesideMutual,pitstop}.

\vspace{1ex}
\noindent\fbox{\begin{minipage}{26em}
\textbf{C3.} Developers have no support for querying multiple microservice database states consistently and they end up encountering challenges on reasoning about the application state.
\end{minipage}}
\vspace{-1ex}

\subsection{Feral ordering}
\label{subsec:feral_ordering}

\noindent \textbf{Description.} Consider an application in the e-commerce domain. After adding several items to a cart, which is managed by the \textit{cart} microservice, the customer may initiate the order's payment process through the API of the \textit{payment} microservice. Suppose that prior to submitting the order, the \textit{cart} microservice emits an event representing the change of the price of one of the items in the customer's cart.

Two options may apply in this case: (i) Application constraints require that events related to changes in the price of items should be reflected in the orders submitted for processing; or (ii) application constraints are such that price changes should not affect such orders. However, we noticed cases where the application does not enforce any constraint. Both options may apply depending on the eventual delivery of events in the system.

\lstset{style=mySharpC}
\begin{lstlisting}[caption={Feral ordering example},label={lst:feral_ordering},captionpos=b]
// Basket microservice: Basket checkout asynchronous request
public async Task CheckoutAsync(Checkout c_out,long userId){
    var basket = await _repository.GetBasketAsync(userId);
    var eventMessage = new CheckoutAcceptedEvent(userId, c_out.Buyer, c_out.RequestId, basket);
    try { _eventBus.Publish(eventMessage); }
    catch (Exception ex){
        _logger.LogError(ex, "ERROR Publishing integration event: {IntegrationEventId}", eventMessage.Id);
        throw;
    }
}
------------------------------------------------------------
// Catalog microservice: Update product asynchronous request
public async Task UpdateProductAsync(Item productToUpdate){
    // code omitted
    if (productPriceChanged){
        var priceChangedEvent = new ProductPriceChangedEvent( 
            catalogItem.Id, productToUpdate.Price, oldPrice);
        await _catalogEventService.
            PublishThroughEventBusAsync(priceChangedEvent);
    }
}
------------------------------------------------------------
// Basket microservice: Reaction to ProductPriceChangedEvent 
public async Task Handle(ProductPriceChangedEvent @event){
    var userIds = _repository.GetUsers();
    foreach (var id in userIds){
        var basket = await _repository.GetBasketAsync(id);
        await UpdatePriceInBasketItems(@event.ProductId, @event.NewPrice, @event.OldPrice, basket);
    }
}
\end{lstlisting}
\vspace{-1ex}

Consider the example adapted from the project eShopContainers \cite{eShopOnContainers} shown in Listing~\ref{lst:feral_ordering}. The \textit{Basket} microservice receives basket checkout requests through the method \textit{CheckoutAsync}. Additionally, the \textit{Catalog} microservice, upon an item price update request from a user, dispatches a \textit{ProductPriceChanged} event to interested parties. As there is no synchronization, whether the new price will be reflected in the user's order depends on the eventual arrival of events, potentially violating application constraints. The same pattern is found in other projects \cite{eShopOnContainers,vertx,LakesideMutual,pitstop,event_stream}.

\noindent \textbf{Discussion.} Literature mentions that developers face challenges in specifying the consistency requirements for their applications~\cite{tanenbaum:2016}. Given that microservices comprise a class of applications that adopt a distributed architectural style~\cite{newman:15}, it is understandable that defining consistency requirements for these applications may become even more challenging. Indeed, there is a lack of support for developers in reasoning about event-based consistency requirements. 


\vspace{1ex}
\noindent\fbox{\begin{minipage}{26em}
\textbf{C4.} Due to the complex interplay between microservices' behaviors, asynchronous events are generated to trigger computations. However, avoiding anomalies related to the unintended interleaving of events across microservices is a challenging task.
\end{minipage}} 

\subsection{Replication hell}
\label{subsec:replication_hell}

\noindent \textbf{Background.} As mentioned in Section~\ref{sec:computations}, microservice architectures may rely on replicating data items across different functional silos to avoid employing synchronous requests spanning multiple microservices for data retrieval and subsequent application-level aggregation. In this context, we observed the prominence of ad-hoc mechanisms for replication. This practice is characterized by the absence of ordering guarantees regarding updates to different objects. The data items arriving from different microservices are often aggregated in queries without any consistency guarantee, i.e., not reflecting a view of the entire system at a single point in time~\cite{ports2010transactional}. This pattern is observed in several microservice applications~\cite{ftgo,eShopOnContainers,LakesideMutual,sentilo,pitstop,event_stream}.

\noindent \textbf{Discussion.} Although one may argue some applications rely on a weak consistency model, such as eventual consistency, to support state querying, it is important to highlight that not all applications fall in this category. Effective mechanisms to support developers are important. In this vein, solutions such as Synapse~\cite{Viennot:15} hold potential to provide more principled replication guarantees in microservices. Even though Synapse~\cite{Viennot:15} is grounded on the industry-strength MVC pattern~\cite{gamma1995design} and meets the polyglot persistence paradigm of microservices (by allowing seamless integration of heterogeneous databases), to the best of our knowledge, the solution has no implementation in frameworks for a variety of programming languages and databases so far, which may hinder a wider industrial adoption. As witnessed in our results, reaching different software development communities is an important feature.

\vspace{1ex}
\noindent\fbox{\begin{minipage}{26em}
\textbf{C5.} The lack of comprehensive support for data replication across microservices lead developers to rely on ad-hoc application-level replication mechanisms, a choice that often leads to inconsistency among microservice states. 
\end{minipage}}

\subsection{Non-transactional queuing}
\label{subsec:queuing}

\noindent \textbf{Background.} In an intra-microservice transaction, updates affecting other microservices are queued for asynchronous processing. As advocated by Pritchett~\cite{pritchett:08}, this queuing must be part of the transactional context of the database operation in the originating microservice. However, as observed in our extended version~\cite{extended}, there is no evidence of use of message persistence queues integrated with the database. Practitioners rely on external message persistence queues (e.g., message brokers) without employing a distributed commitment protocol for message queuing.

Consider the example adapted from the project Lakeside Mutual~\cite{LakesideMutual}, shown in Listing~\ref{lst:non_transactional_queuing}. The method \textit{createInsuranceQuote} performs an insert operation in the database regarding the data item \textit{insuranceQuote}. Afterward, a message representing the operation is queued without a transactional guarantee. In case of error, logging is performed, but no additional measures are taken by the application, which continues its execution. This pattern is identified in several applications~\cite{sentilo,LakesideMutual,pitstop,event_stream}.

\lstset{style=myJava}
\begin{lstlisting}[caption={Non-transactional queuing example},label={lst:non_transactional_queuing},captionpos=b]
public ResponseEntity createInsuranceQuote() { 
	InsuranceQuote insuranceQuote = new InsuranceQuote
	    (date, QUOTE_SUBMITTED, insuranceOptionsEntity);
	insuranceQuoteRepository.save(insuranceQuote);
	policyMessageProducer.send(date, insuranceQuote);
	return ResponseEntity.ok(insuranceQuote);
} // parameters omitted
------------------------------------------------------------
public void send(Date date, InsuranceQuote insuranceQuote) {
    InsuranceQuoteEvent insuranceQuoteEvent = 
	    new InsuranceQuoteEvent(date, insuranceQuote);
	try { jmsTemplate.send(insuranceQuoteEventQueue, insuranceQuoteEvent);
	} catch(JmsException exception) {
		logger.error("Failed to send", exception);
	}
}
\end{lstlisting}
\vspace{-1ex}

\noindent \textbf{Discussion.} Existing database systems often provide support for persistent messaging or notification mechanisms integrated into the database~\cite{postgresql_notify}. However, to use these mechanisms, different clients are forced to access the same database instance. As a result, most open-source projects and the literature rely on external message brokers for the communication and event exchange between microservices. Without interoperability of databases and message brokers, 2PC is not possible, violating the atomicity property prescribed by Pritchett~\cite{pritchett:08}.

\vspace{1ex}
\noindent\fbox{\begin{minipage}{26em}
\textbf{C6.} Developers lack viable and efficient abstractions for transactional queuing in microservice architectures. As a result, anomalies arise due to lack of isolation and ad-hoc fault-handling, leading to challenges on ensuring application correctness. 
\end{minipage}}
\vspace{-1ex}



\section{Delving into Developers' Pains}
\label{sec:pains}


To strengthen our confidence on the practical challenges revealed  in the last section 
and derive additional ones,
we inquired practitioners about pressing challenges they face 
while dealing with data management in microservices.

We present next an aggregated discussion over the responses collected from the participants. When appropriate, we include the percentage of each response and the challenges associated. Details about the questions and overall methodology followed in this questionnaire can be found in our extended version~\cite{extended}.


\noindent \textbf{A. Constraint enforcement and schema evolution.} In line with \textbf{C1} and \textbf{C2},
fixing data inconsistencies (19\%) and enforcing correctness through application-level validations (24.5\%) are prevalent challenges, which, in conjunction with descriptions provided by the respondents, revealed two major interrelated issues.

On the one hand, application-level data consistency and integrity problems, such as ``app[lication] code was not [...] removing all usages of an item (in a NoSQL DB[MS])'' and ``ensur[ing] data is persisted correctly and not duplicated'' are mentioned by participants. On the other hand, the use of asynchronous communication for cross-microservice operations or for data replication introduces challenges when it comes to schema evolution in individual microservices, matching our hypothesis highlighted in \textbf{O2}. For instance, one respondent declared that ``in an async[hronous] environment and having microservices be[ing] responsible for processing their own changes, issues introduced with new releases on microservices may cause inconsistencies in data processing which are generally hard to correct after the fact.'' Another respondent declared that ``as there is no one ruling constraint system, data cleanliness is a significant challenge.''

\vspace{1ex}
\noindent\fbox{\begin{minipage}{26em}
\textbf{C7.} Changes made to a microservice's schema might necessitate adaptations in the structure of messages exchanged; however, application logic in dependent microservices could still be making conflicting assumptions regarding invariants.
\end{minipage}}
\vspace{1ex}



\noindent \textbf{B. Eventual consistency.} In line with open-source projects and the literature, respondents have also pointed out eventual consistency (15.68\%) as a challenge. A participant declared: ``Dealing with eventual consistency was particularly hard when convergence happens into a broker state. The complexity of the approaches hands a great barrier for developers.'' We observed two primary drivers for non-converged state in microservices: (i) resorting to event-driven and asynchronous replication to avoid synchronous communication and consequently high latency in online queries (Section~\ref{subsec:replication_hell}); (ii) employing workflow-oriented business transactions across microservices, often driven by asynchronous and event-based communication (Section \ref{sec:computations}).

In both cases, reasoning about the global state of the application is a major concern. In the first case, it is hard to determine when replication has occurred to a sufficient degree, especially as ad-hoc mechanisms are often employed by practitioners. In the second, it is challenging to assert whether a multi-microservice workflow has terminated and what its current state is.

\vspace{1ex}
\noindent\fbox{\begin{minipage}{26em}
\textbf{C8.} Due to the distributed nature of microservice architectures, eventual consistency is often taken as the de facto consistency model by practitioners. This choice introduces a series of challenges on reasoning about distributed states and invariants.
\end{minipage}} 
\vspace{1ex}



\noindent \textbf{C. Ensuring consistency between heterogeneous database systems.} Although one may argue that the eventual consistency approach, i.e., convergence through asynchronous events representing data item updates, is a sufficient consistency model for most microservice-based applications, our survey results show that there is a class of applications that requires stronger guarantees over writes performed in different database systems (14.70\%). For example, a participant argued that ``atomicity can not be guaranteed over different storage technologies, no information or proper literature. Guessing and fixing error approach.''

In addition, the prevalence of in-memory caching systems for increasing throughput and decreasing data access latency in microservices ($\S$~\ref{subsec:database_patterns}) may also introduce challenges when developers resort to asynchronous writes. For instance, a respondent declared: ``writes are asynchronous to remove the DB[MS] from the critical path [of a data processing pipeline] and ensure that messages can be delivered in near-realtime, [...] but we've already had few situations where the cache expired before the information had actually had time to be persisted in the DB,'' suggesting measures outside the database were necessary to correct the anomaly.

\vspace{1ex}
\noindent\fbox{\begin{minipage}{26em}
\textbf{C9.} Developers deal with data inconsistencies with a myriad of ad-hoc measures, such as manual intervention to the underlying microservice databases as well as applying custom-built data management logic to handle possible inconsistencies.
\end{minipage}}
\vspace{1ex}


\noindent \textbf{D. Consistent querying.} As revealed before ($\S$~\ref{subsubsec:online_querying} and \textbf{C3}), online queries are prominent in microservice architectures. However, there is no de facto approach as developers tend to rely on a variety of ad-hoc mechanisms for online queries (8.82\%). Moreover, a consistent view of global state is also expressed as a requirement: ``We are integrating data from different sources in a global transportation network. The changes in data are flowing into our system consistently. We need to integrate as fast as possible to present a `picture of the moment' to the global end-users.'' To this matter, a respondent indicated ``polystore databases [as a solution] to provide location independence and semantic completeness for queries performed by heterogeneous microservices.''

\noindent \textbf{E. Poor functional partitioning.} Some participants indicated poor functional partitioning as a potential cause of data consistency problems. For instance, one argued that ``microservices made people separate code when they should not be separated, causing this eventual consistency everywhere. [...] people wanted to create a separate microservice, just because of `size', and we end up having consistency problems.'' Although initial work suggests considering database-related attributes, such as relationships between relations to derive a functional partitioning~\cite{LaignerLKPS19:ReactorPart}, further explorations with real-world systems are necessary.




\vspace{1ex}
\noindent\fbox{\begin{minipage}{26em}
\textbf{C10.} Decomposing application functionality without proper thought onto constraints and consistency requirements may lead to the burden of dealing with data inconsistencies.
\end{minipage}}
\vspace{1ex}



\noindent \textbf{F. Limitations of benchmarks.} We realized that existing benchmarks (e.g., DeathStarBench ~\cite{gan2019open}) fail to incorporate the asynchronous and event-driven properties of microservices. This can be challenging for programmers to test new solutions. 

\vspace{1ex}
\noindent\fbox{\begin{minipage}{26em}
\textbf{C11.} The lack of a benchmark that properly reflects real-world deployments refrains developers from effectively experimenting with and reasoning about microservice deployments.
\end{minipage}}
\vspace{1ex}


\noindent\textbf{Summary.} The perception of developers regarding the challenges faced while dealing with data management are very aligned with our findings discussed in $\S$~\ref{sec:microservices_glass}. Additional challenges were revealed, such as schema evolution (\textbf{C7}) and data cleaning (\textbf{C9}). This overall perception strengthens our confidence that there is an unaddressed need for effective data management support in microservices.


\section{Towards Microservice-oriented Database Systems}
\label{sec:towards}

In this section, we devise a candidate set of features for a microservice-oriented DBMS based on the observations and challenges identified in the previous sections. Then, we turn our attention to explain why state-of-the-art DBMSs are insufficient for the needs of microservices in the light of the proposed features. Finally, we discuss how to realize the features into a microservice-oriented DBMS.

\vspace{1ex}

\noindent\textbf{What features are required for a microservice-oriented DBMS?}



As explained in $\S$~\ref{subsec:feral_ordering}, developers have no way to specify event-based constraints. For example, concurrent checkout requests in eShopContainers~ \cite{eShopOnContainers} may be processed in arbitrary order wrt. other events, hence providing no guarantee of applying updated prices before checkout. 

\vspace{1ex}
\noindent\fbox{\begin{minipage}{26em}
\textbf{F1.} The database system should be aware of events cutting across several microservices, so that it can capture the complex interplay of data exchange between microservices. 
\end{minipage}}
\vspace{0.5ex}

\vspace{0.5ex}
\noindent\fbox{\begin{minipage}{26em}
\textbf{F2.} Event-based awareness allows the database system to enforce event-based constraints, so that application safety can be supported. 
\end{minipage}}
\vspace{1ex}

Besides, as explained in $\S$~\ref{subsec:motivations}, $\S$~\ref{subsec:business_transactions}, and $\S$~\ref{subsubsec:event_driven_computing}, 
practitioners employ communication through asynchronous events to decouple microservices. However, it is observed that, by decoupling, developers have a hard time enforcing high data consistency ($\S$~\ref{subsec:feral_ordering}, $\S$~\ref{subsec:replication_hell},$\S$~\ref{subsec:queuing}).





\vspace{1ex}
\noindent\fbox{\begin{minipage}{26em}
\textbf{F3.} The database system should provide abstractions that strike a balance between loose coupling among components and high data consistency, so that effective and efficient cross-microservice coordination can be achieved. 
\end{minipage}}
\vspace{1ex}



Constraints cutting across several microservices are often enforced through application-level validations ($\S$~\ref{subsec:cross_validations}). As explained in $\S$~\ref{subsec:implicit}, 
such an approach leads to an implicit logical data dependency -- given that it is not formalized as a constraint -- from the \textit{Cart} microservice to the \textit{Product} microservice.


\vspace{1ex}
\noindent\fbox{\begin{minipage}{26em}
\textbf{F4.} The database system should allow users to specify cross-microservice associations and enforce them consistently, so that users can avoid encoding error-prone validations at the application level. 
\end{minipage}}
\vspace{1ex}



In addition, as microservices extensively employ cache stores to reduce the costs of querying database states, the constraints across microservices should also take into account the consistency of the data stored in the underlying microservices' cache stores (\textbf{C9}).

\vspace{1ex}
\noindent\fbox{\begin{minipage}{26em}
\textbf{F5.} Database systems should allow the user to specify the freshness semantics between the underlying cache and database states, so that application safety can be safeguarded through adequate cache consistency management at the microservice level. 
\end{minipage}}
\vspace{1ex}



As explained in $\S$~\ref{subsubsec:online_querying}, replication is often employed so that microservices can avoid coordination via cross-microservice queries. Microservices then queue or publish their own data items' updates as asynchronous events, which are then eventually delivered to consumer microservices. However, expressing data replication through application-level mechanisms creates a great barrier for developers.


\vspace{1ex}
\noindent\fbox{\begin{minipage}{26em}
\textbf{F6.} The database system should offer effective abstractions for users to specify data replication across microservices, so that varied replication consistency models can be properly supported.
\end{minipage}}
\vspace{1ex}



In $\S$~\ref{subsubsec:online_querying} and $\S$~\ref{sec:pains}, we observed the practice of retrieving data from several microservices to build a real-time view. However, microservice developers find challenges when implementing consistent cross-microservice queries, such as point-in-time views ($\S$~\ref{subsec:cross_queries}).


\vspace{1ex}
\noindent\fbox{\begin{minipage}{26em}
\textbf{F7.} The database system should provide native support for consistent cross-microservice online queries, but at the same time respect the data sovereignity principle of microservices.
\end{minipage}}
\vspace{1ex}

\noindent\fbox{\begin{minipage}{26em}
\textbf{F8.} The database system could optimize cross-microservice queries through pre-processing the data, pre-joining the data, or proactively replicating the data owned by different microservices while fulfilling the consistency-isolation levels required by the application.
\end{minipage}}
\vspace{1ex}

\noindent\textbf{Why are Web-Scale Application Architectures not Enough?}
\label{subsub:web_scale_not_enough}

Although microservice applications can be considered a realization of BASE, an approach where functional partitions are eventually consistent, the microservice paradigm introduces data management requirements not originally envisioned by Pritchett~\cite{pritchett:08}, including but not limited to application-level replication, consistent cross-microservice queries, cross-microservice validations, heterogeneous services, events cutting across several microservices, and interleaving of distinct, but correlated events that lead to data races. As prescribed by the BASE model, functional partitioning necessarily involves pushing constraint enforcement to the application level~\cite{pritchett:08}, which as we have seen often leads to consistency problems in microservice architectures ($\S$~\ref{sec:microservices_glass}). Lastly, BASE can be considered a set of principles for designing functionally partitioned applications, thus not a full-fledged data management solution.

Furthermore, there have been several recent works in Internet-scale database services. These systems provide high availability at a global scale, at the same time offering high throughput data processing and multi-tenancy by design. While some of them support strong consistency guarantees, such as Google Spanner~\cite{spanner}, Amazon Aurora~\cite{aurora}, and Azure Cosmos DB~\cite{cosmos_db}, others trade stronger consistency for performance, such as DynamoDB~\cite{dynamo} and Astra DB~\cite{astra_db}. We address the common shortcomings of these solutions in the context of the aforementioned features.






\noindent\textbf{F1 \& F2.} In addition to the lack of atomic event queuing in Internet-scale databases ($\S$~\ref{subsec:queuing}), database systems are unaware of the complex interrelationships among events in microservice applications. Thus, it would be desirable for developers to be able to specify event-based constraints on their processing, with the enforcement being performed at the database layer ($\S$~\ref{subsec:feral_ordering}).




Consider, for example, a scenario found in popular e-commerce microservice applications~\cite{eShopOnContainers,vertx,sockshop,event_stream} where an order checkout is optimistically processed, i.e., the products in the order are sold without checking their availability in stock. Connecting with Figure~\ref{fig:diff_mono_msa}(b), the \textit{Stock} microservice might subscribe to checkout and payment events in order to determine whether the process of acquisition of new stock should be triggered. However, the \textit{Stock} microservice should only take into account checkout events whose corresponding payment events have arrived, since the sale of the products is only finally confirmed after payment.


In the presented scenario, resorting to a single tenant in a multi-tenant database system would allow practitioners to write database triggers for the latter functionality; however, this approach introduces an undesirable dependence between the schemas of the various microservices. By contrast, practitioners could resort to multiple tenants, e.g., one per microservice for isolation. Unfortunately, in this case, the checking of the stock replenishment condition would need to be pushed to the application level.

\noindent\textbf{F3.}
As pointed out above, it is not obvious how to employ a multi-tenant DBMS to store the states of distinct microservices. One solution is to employ one tenant per microservice, but this leads the states to be completely isolated. Another possibility is to use a single tenant for the states of multiple microservices; however, the latter implies complete sharing. For instance, Figure~\ref{fig:diff_mono_msa}(b) depicts the case where Orders requires coordinating with Stock, Campaigns, and Discounts to safeguard that an order transaction is correctly placed. Coordinating these 4 microservices would require either resorting to application-level 2PC or ad-hoc application-level coordination mechanisms (\textbf{O5}, \textbf{O6}) -- in case states were partitioned across tenants -- or use of DBMS transactions -- in case states were shared. In either case, the principles of loose-coupling and autonomy among microservices would be jeopardized. On the one hand, application-level 2PC would imply API coupling with high consistency, while ad-hoc application-level coordination would typically entail loose coupling at lower data consistency levels. On the other hand, sharing of the same database tenant would imply that microservices lose the ability to independently evolve and fail.

\noindent\textbf{F4 \& F5 \& F6.} In the same line of reasoning of \textbf{F3},  
multi-tenant database systems (i) cannot enforce cross-microservice constraints, (ii) cannot reason about cache consistency over distinct microservices' database states, and (iii) cannot provide consistent data replication across tenants without failing to meet the decentralized data management and data sovereignity principles prescribed by the microservice paradigm. Consider, for example, the case where the state of \textit{Payment} microservice refers to the customers relation in the state of the \textit{Customer} microservice. By resorting to a single tenant for both microservices, we could potentially express a foreign key across their respective schemas. However, that would create a schema and failure dependency between the microservices. 
By resorting to different tenants, there would be no way to express the foreign key constraint across the two microservice states. 

\noindent\textbf{F7 \& F8.} 
Consider a scenario where in the context of a customer's complaint about an undelivered package, a front-end microservice may need to provide a real-time view over the customer's interactions, discounts being applied, and the customer's credit score to enable business users to decide whether the customer can receive an additional compensation for what happened or only be reimbursed. Such an operation necessitates querying data from several microservices' states. However, it is unclear how existing database systems can support consistent cross-microservice queries. 

Simply providing querying services over the private states of multiple tenants is insufficient to achieve consistent views. Rather, the DBMS needs to capture the interactions between the microservices so that it can analyze and support consistency semantics.
Although a DBaaS vendor could potentially provide a holistic distributed state view across tenants (considering all microservice states are deployed as a tenant), such a feature would still be insufficient if not addressed in conjunction with all the prescribed features that necessarily require knowing more about the application, as we address next.


 
\noindent\textbf{How to realize the features into a microservice-oriented DBMS?}

Usually, data management tasks executed at the application level are a black-box to the database system~\cite{powerstation}. As a result, the database system is often unaware of the complex data management logic and constraints put forth by application developers~\cite{bailis:15}. This impedance mismatch is particularly worsened when it comes to microservices, since data is flowing outside of the database~\cite{data_outside}, that is, there are many data management tasks that cut across microservices. The DBMS is oblivious to these tasks. To realize the features described, we believe 
it is essential for the database system to better understand what is occurring at the application level. To bridge this gap, we envision two paths:

\noindent(A) Extending current abstractions for building stateful middle-tier applications can be seen as a fruitful solution. Laudable initial efforts on virtual actors~\cite{bernsteinorleans} and stateful functions~\cite{stateful_functions} have not yet incorporated the features required by microservice applications to a sufficient degree. For example, in Orleans, data durability functionality is explicitly managed by the developer. In addition, there is no way to specify event-based constraints and it is unknown whether all microservice applications can be modeled through virtual actors~\cite{orleans_best_practices,wang2019modeling};


\noindent(B) Extending the frontiers of a database system by 
having database components living in the application-tier and communicating the application's needs to the database system. 
With current database APIs, applications have no way to push down complex dataflows to the database system in a meaningful way, because interfaces abstract away the complex interaction and data exchanges happening outside the database. Although an initial proposition in this line has been presented in ~\cite{virtual-ms}, no formalism and implementation has been provided so far.

\vspace{1ex}
\noindent\fbox{\begin{minipage}{26em}
\textbf{F9.} Appropriate abstractions should be formalized to allow the database system to gain knowledge of the data management logic carried out by the application and the complex interplay of microservices. These abstractions could help in achieving a proper division of tasks between the application and the database system, where data management tasks are pushed to the database system, thus alleviating the burden on developers.
\end{minipage}}
\vspace{1ex}
\section{Conclusion}
\label{sec:conclusion}






In this paper, we observe that the lack of a holistic data management solution for microservices leads practitioners to resort to \textit{ad-hoc} designs. These are characterized by weaving together heterogeneous services and resorting to application-level data management mechanisms,
 which lead to problems that cannot be resolved through code refactorings~\cite{fowler:99}. To tackle the data management requirements of microservices by design, we present a set of candidate features for database systems so that they can play a central role in this new paradigm.

\begin{acks}


This project has received funding from the European Union's Horizon 2020 research and innovation programme under the Marie Skłodowska-Curie agreement No 801199 and Independent Research Fund Denmark grant No 9041-00368B.

\if 1\preprint
We are grateful to the practitioners that collaborated to this work, gently providing their insights. We are particularly grateful to Diogo Souza (Guiabolso), Gabriel Lima (VTEX), Leonardo Gomes (Amadeus), and Pedro Diniz (VTEX) for their valuable comments. 
\fi

\noindent Dedicated to the memory of Maria José Teixeira $\dagger$.

\end{acks}


\bibliographystyle{ACM-Reference-Format}
\bibliography{sample}

\if 1\preprint
\appendix
\input{appendix/09_appendix}
\fi

\end{document}